\documentclass[showpacs,twocolumn,prb]{revtex4}
\usepackage{graphicx, amsmath, amsthm, amssymb}
\newcommand{\be}{\begin{equation}}
\newcommand{\ee}{\end{equation}}
\newcommand{\bea}{\begin{eqnarray}}
\newcommand{\eea}{\end{eqnarray}}
\newcommand{\bwt}{\begin{widetext}}
\newcommand{\ewt}{\end{widetext}}

\newcommand{\up}{\uparrow}
\newcommand{\dn}{\downarrow}

\begin{document}

\title{Two-site dynamical mean field theory for the dynamic Hubbard model}
\author{G.H. Bach}
\affiliation{Department of Physics, University of Alberta, Edmonton, AB, Canada T6G~2G7}
\author{J.E. Hirsch}
\affiliation{Department of Physics, University of California, San Diego, La Jolla, CA 92093-0319}
\author{F. Marsiglio}
\affiliation{Department of Physics, University of Alberta, Edmonton, AB, Canada T6G 2G7}

\begin{abstract}At zero temperature, two-site dynamical mean field theory is applied to the Dynamic Hubbard model. The Dynamic Hubbard model describes the orbital relaxation that occurs when two electrons occupy the same site,
by using a two-level boson field at each site. At finite boson frequency, the appearance of a Mott gap is found to be enhanced even though it shows a metallic phase with the same bare on-site interaction $U$ in the conventional Hubbard model. The lack of electron-hole symmetry is highlighted through the quasi-particle weight and the single particle density of states at different fillings, which qualitatively differentiates the dynamic Hubbard model from other conventional Hubbard-like models.

\end{abstract}


\date{\today}

\maketitle

\section{Introduction}
Dynamic Hubbard models \cite{hirsch01} represent a new class of model Hamiltonians that describe the modification of the electronic wavefunction that occurs when an atomic orbital becomes
occupied by more than one electron.  The key difference between  dynamic and standard Hubbard-like models can be understood by considering a Helium atom.  Helium has two electrons in the $1s$ shell and the strength of the interaction between them is comparable to the electron-ion interaction. As a consequence, the two-electron wave function cannot be simply represented by  a single Slater determinant
formed by the electronic wavefunction of the singly occupied orbital ($He^+$) for each electron, since the wavefunction of one electron is modified by the presence of the
other  electron.  The effect is more pronounced in the negative ion $H^-$ because of the weaker attraction between the electrons and the nucleus.
Conventional tight binding models like the Hubbard model completely neglect this effect since they
assume that the wavefunction for two electrons on a site is a simple product of the wavefunctions for a singly occupied site\cite{hirsch01}.

The Dynamic Hubbard model tries to include the physics of orbital relaxation  by modulating the on-site interaction term (Hubbard $U$) with an auxiliary boson (or spin) degree of freedom.\cite{hirsch01} An effective low-energy Hamiltonian for this model modifies the hopping term so that the hopping amplitude depends on the electronic occupation of the sites involved in the hopping process;
 this is known as the correlated hopping model \cite{hirsch90}.
In this paper, we focus on the dynamic Hubbard model with an auxiliary spin-$1/2$ degree of freedom; the two states of this pseudospin can be viewed as describing the
modification of the electronic wavefunction upon double occupancy. This kind of model was first suggested two decades ago.\cite{hirsch89_pla} The model was studied further through world line quantum Monte Carlo (QMC) methods \cite{hirsch02b}, exact diagonalization (ED) \cite{hirsch02c}, and an approximate perturbative analysis starting from a generalized Lang-Firsov transformation.\cite{marsiglio03} The exact treatments suffer from small size effects, but more recently, Bouadim et al.\cite{bouadim08} also studied this model using determinant QMC on somewhat larger clusters ($N=6\times6$); however, due to the fermion sign problem, they encountered difficulties, especially at low temperatures.

In recent years it has become clear that dynamical mean field theory (DMFT) is a valuable way to treat the local aspects of both quasiparticles and incoherent high energy excitations on the same footing in strongly correlated electron systems.\cite{georges96} DMFT is practical because, instead of considering a large lattice model whose Hilbert space is exponentially large, one needs to solve merely a single impurity model; for this problem various algorithms for an exact solution already exist. \cite{hirsch86} Even though the single impurity model can be treated numerically by QMC, it is still a computationally expensive problem. An approximate but effective alternative was proposed by Potthoff, \cite{potthoff01} who proposed the so-called ``two-site DMFT''; this method uses two sites, one for the impurity and one for the `bath' of conduction electrons, which is readily solved exactly. Using only one bath site renders the mapping to the single impurity Anderson model (SIAM) approximate, but the self-consistency is now easily controlled through the bath parameters. In Ref. \onlinecite{potthoff01} the validity of the two-site DMFT simplification was established for the Hubbard model; for example, it predicts qualitatively correctly the existence of a Mott transition critical point, and even though it is not a conserving theory in the sense of Baym and Kadanoff, \cite{baym61} the violation of the Luttinger sum rule, for example, is fairly small. Given our interest in establishing qualitative trends for the Dynamic Hubbard model, and the existence of QMC results (albeit for small lattices) as a benchmark, we will adopt the two-site DMFT approximation to study the dynamic Hubbard model.

There have been some studies of correlated hopping models using DMFT, in particular applied to the Falicov-Kimball model\cite{fk1,fk2}. Because the interaction term involves two sites rather than one, the self-energy is  non-local  and
the formalism becomes considerably more complicated. Instead with the dynamic Hubbard model considered here
we can describe the physics of correlated
hopping in a much simpler way with a single-site self-energy, by simply considering the model in the limiting case where the
interaction becomes non-retarded (large $\omega_0$ in the Hamiltonian Eq. (1)).

As we illustrate below, the two-site DMFT treatment of the Dynamic Hubbard model gives  semi-quantitative agreement with the QMC results. For half-filling and below, the properties of the Dynamic Hubbard model mimic those of the Hubbard model. For example a Mott insulating phase appears for strong enough on-site interaction, and, with an attractive on-site interaction, pairing is enhanced. In addition, however, electron-hole asymmetry naturally arises; this is evident, for example, in the dependence of the quasi-particle weight on the electron/hole number density. Thus this model captures the essential physics that a few electrons in a nearly empty electronic energy band can behave very differently from a few holes in a nearly filled band.

The paper is organized as follows: The next section will briefly describe the Dynamic Hubbard model  with an auxiliary spin $1/2$ degree of freedom and will provide a synopsis of the two-site DMFT approximation. In section III we present some numerical results and discuss some of the characteristic properties of the dynamic Hubbard model, especially those that differentiate it from the simple Hubbard model. In addition, we show some comparisons with the QMC results, \cite{bouadim08} which indicate that the two-site approximation works very well. The last section IV will summarize our results and suggest directions for further study.

\section{Model and Method}

We consider here the Dynamic Hubbard Hamiltonian \cite{hirsch01} with a spin-$1/2$ degree of freedom in the electron representation:
\bea
H_{\rm DHM}= \sum_{<i,j>\sigma}t_{ij}(c^{\dagger}_{i\sigma}c_{j\sigma}+c^{\dagger}_{j\sigma}c_{i\sigma}) - \mu \sum_{i,\sigma}n_{i\sigma} \nonumber \\
+ \sum_{i}(\omega_{0}\sigma^{x}_{i}+ g\omega_{0}\sigma^{z}_{i}) + \sum_{i}(U-2g\omega_{0}\sigma^{z}_{i})n_{i\up}n_{i\dn}.
\label{ham_dhm}
\eea
The first term is the electron hopping term; $c^{\dagger}_{i\sigma}$ ($c_{i\sigma}$) is an electron creation (annihilation) operator at site $i$ with spin $\sigma$. Following Potthoff \cite{potthoff01} we use a Bethe lattice with infinite connectivity with nearest neighbour hopping only, so that $t_{ij}= -t < 0$ for nearest neighbours only. The parameter $t = t^\ast/\sqrt{q}$, with $q$ the connectivity, and $t^\ast = 1$ sets the energy scale. The second term is the usual chemical potential term which determines the electron filling. The auxiliary spin degree of freedom is given in the third term; the two levels have a spacing given by $\omega_0$. The fourth term describes interactions between two electrons. In addition to the onsite Hubbard U term, there is an additional coupling to the auxiliary spin degree of freedom. As explained in Ref. \onlinecite{hirsch01} and reviewed in the next section, this term varies the actual on-site repulsion, dependent on the state of the auxiliary degree of freedom.

Dynamical Mean Field Theory has been widely studied in a number of correlated fermion systems, and this approach has been quite successful, as reviewed in Ref. \onlinecite{georges96}. In particular, in models where the {\em local dynamics} is expected to play the most important role (as opposed to spatial correlations), the DMFT without the use of cluster methods \cite{maier05, potthoff05} should be accurate. The Dynamic Hubbard model should be ideally suited for these conditions. DMFT maps a lattice model onto a quantum single impurity model through self-consistent conditions; in this paper, we consider particularly the single impurity Anderson model (SIAM):
\bwt
\bea
H_{imp} = \sum_{\sigma} (\epsilon_{d}-\mu)d^{\dagger}_{\sigma}d_{\sigma} + \sum^{n_{s}}_{\sigma,k=2} (\epsilon_{k} -\mu)a^{\dagger}_{k\sigma}a_{k\sigma}
 + \sum^{n_{s}}_{\sigma,k=2} V_{k}(d^{\dagger}_{\sigma} a_{k\sigma}+h.c.)
 + (\omega_{0}\sigma^{x}+ g\omega_{0}\sigma^{z}) + (U-2g\omega_{0}\sigma^{z})n_{d\up}n_{d\dn}
\eea
\ewt
where $d_\sigma$, $d^{\dagger}_\sigma$ are the impurity operators with spin $\sigma$ which only act on the single impurity orbital with energy $\epsilon_{d}$. This impurity site is hybridized to a bath with $n_s-1$ degrees of freedom through the coupling parameter $V_k$; the Coulomb interaction $(U-2g\omega_{0}\sigma^{z})$  only occurs when two electrons are on the impurity site. The on-site Green's function for the lattice model is given by
 \be
 G(\omega) = \int^{\infty}_{-\infty} dx \frac{\rho(x)}{\omega+\mu-\Sigma(\omega)-x}
\label{gw}
 \ee
 where $\Sigma(w)$ is the local self-energy in infinite dimensions ($q \rightarrow \infty$) and $\rho(x)$ is the free density of electron states for a Bethe lattice: \\
 \be
 \rho(x) = \frac{1}{2 \pi {t^{\star}}^{2}} \sqrt {4{t^{\star}}^{2}-x^{2}}.
 \label{dense0}
 \ee
 Note that a `momentum' dependent spectral Green function is implied by Eq. (\ref{gw}), and, for momentum corresponding to bare energy $x$, is given by
 \be
 G(x,\omega)^{-1} \equiv {\omega + \mu - \Sigma(\omega) - x};
 \label{gwx}
 \ee
 this implies a spectral function $A(x,\omega) \equiv -{1 \over \pi} {\rm Im} G(x,\omega)$, to be used later in the optical conductivity.

 Because of the simplicity of the non-interacting density of states, Eq. (\ref{gw}) can be integrated analytically. The result is:
 \be
 G(\omega) = {1 \over t}\biggl( {\omega + \mu - \Sigma(\omega) \over 2t} - \sqrt{\bigl({\omega + \mu - \Sigma(\omega) \over 2t}\bigr)^2 - 1}\biggr),
 \label{gwanal}
 \ee
where the $\sqrt{()}$ is taken with a sign equal to ${\rm sgn} [{\rm Re}(\omega + \mu - \Sigma(\omega))]$.

 For the SIAM, the impurity Green's function can be written as:
 \be
 G_{imp}(\omega) = \frac{1}{\omega+\mu-\epsilon_{d}-\Delta(\omega)-\Sigma_{imp}(\omega)}
 \ee
  in which $\Delta(\omega)=\sum^{n_s}_{k=2} V^{2}_{k}/(\omega+\mu-\epsilon_{k})$ is the hybridization function and $\Sigma_{imp}$ is the impurity self-energy. The self-consistent process is based on the local nature of the quantum system in the limit of infinite dimensions, in which the on-site Green's function for the lattice model can be averaged over all momenta and only depends on the frequency as we obtain in Eq. (\ref{gw}). Instead of directly solving the dynamic Hubbard lattice model, we only need to solve the SIAM (with fewer degrees of freedom); from this we obtain the impurity Green's function which should be the same as the on-site Green's function for the lattice:
  \bea
  G(\omega)=G_{imp}(\omega) & \label{gself}\\
  \Sigma(\omega) = \Sigma_{imp}(\omega) \label{sigself}
  \eea
Eqs. (\ref{gself},\ref{sigself})  are the self-consistency conditions.

A solution is required for the impurity Green's function. While an exact solution is available through QMC techniques, \cite{hirsch86} Potthoff suggested a much faster though approximate procedure known as the two-site DMFT, \cite{potthoff01} which he benchmarked for the Hubbard model. In this approach, \cite{potthoff01} the SIAM with only two sites is diagonalized; one site is for the impurity and one site represents the bath, so $n_s=2$. Therefore the bath parameters are single numbers, as $\epsilon_{k}=E_{c}$ and $V_{k}=V$. Besides making the procedure significantly faster, the two-site DMFT is more transparent, as the self-consistency conditions are analytic. The self-consistency conditions (\ref{gself},\ref{sigself}) are replaced by conditions at high and low frequency, to give two new self-consistentcy conditions which relate directly to the bath parameters:
\bea
n_{imp} &=&  n
\label{nself} \\
V^2 &=& zM^{(0)}_{2}
\label{vself}
\eea
where $n_{imp}$ ($n$) is the filling for the impurity site (conduction band in the lattice model). The parameter $z = 1/(1-Re(d\Sigma(0))/d\omega)$ has the meaning of the quasi-particle weight in the metal phase and $M^{(0)}_{2}=\sum_{<i,j>}t^{2}_{ij}$ is the second moment of the non-interacting density of states. For the model adopted here, Eq. (\ref{dense0}), this becomes $M^{(0)}_{2}={t^\ast}^2 \equiv 1$.
Therefore the right-hand-side of Eq. (\ref{vself}) reduces to the quasiparticle weight. In fact, the procedure follows closely that given in Ref. \onlinecite{potthoff01}, and the reader is referred to that publication for full details. Once the impurity problem is solved, one can obtain the density of states for the original lattice through
\be
A(\omega) = -{1 \over \pi} {\rm Im} G(\omega+i0^{+}) = \rho(\omega+\mu-Re(\Sigma(\omega)))
\ee
where the second equality only follows because the self energy is given by a two-pole approximation;\cite{potthoff01} the self-energy is obtained from the self-consistent condition, Eq. (\ref{sigself}) and Dyson's equation,
\be
\Sigma_{\rm imp}(\omega+i0^+) = {G^{(0)}_{\rm imp}(\omega+i0^+)}^{-1} - {G_{\rm imp}(\omega+i0^+)}^{-1}.
\ee
We will also be interested in the behavior of the optical conductivity, which within the same local approximation is given by the expression\cite{georges96}
 \begin{eqnarray}
\sigma_1(\omega)&=&
{e^2t^2a^2 \over \pi \hbar^2 \nu}
\int_{-\infty}^{+\infty} d\epsilon \frac{f(\epsilon)-f(\epsilon+\omega)}{\omega} \times \nonumber \\
& &\int _{-\infty}^{+\infty} dy \rho(y) A(y,\epsilon)A(y,\epsilon+\omega),
\label{cond}
 \end{eqnarray}
 where $a$ is the lattice constant and $\nu = a^d$ is the volume of the unit cell in $d$ dimensions. As stated above,
 the single particle spectral function, $A(x,\omega)$, is defined as
 \be
 A(x,\omega) \equiv -\frac{1}{\pi}{\rm Im} \biggl(\frac{1}{\omega + \mu - \Sigma(\omega) - x}\biggr).
 \label{axw}
 \ee
This function is immediately known once the self-energy is determined.

\section{Results and discussion}

\subsection{Site Hamiltonian}

The Dynamic  Hubbard model Hamiltonian, Eq. (\ref{ham_dhm}), consists of electron degrees of freedom that can move throughout the lattice, along with pseudospin degrees of freedom that reside at each lattice site. The pseudospins model the ability of the ions to `react' to the different
electronic configurations by changing the orbitals when electrons are and are not present. Following Ref. \onlinecite{hirsch02b}, we focus on the on-site Hamiltonian {\em for electrons}:
\be
H^{(i)}_{DHM} = \omega_{0}\sigma^{i}_{x} + g\omega_{0}\sigma^{i}_{z} +[U-2g\omega_{0}\sigma^{i}_{z}]n_{i\up}n_{i\dn},
\label{ham_onsite}
\ee
which is easily solved, given the presence of 0, 1, or 2 electrons. Using the spin-$1/2$ $\sigma_z$ eigenstates, $|+>$,$|->$, as a basis, we find, that with $n$ electrons present, the eigenstates are
\bea
|n>_a&=&u(n)|+>+v(n)|-> \\
|n>_b&=&v(n)|+>+u(n)|->
\label{eigenstates}
\eea
with eigenvalues:
\bea
\epsilon(n)_a &=& \delta_{n,2}U -\omega_{0}\sqrt{1+g^2}
\nonumber \\
\epsilon(n)_b &=& \delta_{n,2}U +\omega_{0}\sqrt{1+g^2}
\label{eigenvalues}
\eea
The eigenvector components are given by
\bea
u(0)&=&u(1)=v(2)\\
v(0)&=&v(1)=u(2),
\label{comp}
\eea
with
\bea
u^2(0)=\frac{1}{2}(1-\frac{g}{\sqrt{1+g^2}}) \\
v^2(0)=\frac{1}{2}(1+\frac{g}{\sqrt{1+g^2}}).
\label{uandv}
\eea
The expectation value of the pseudospin, in the ground state, illustrates the relaxation of the orbital required, depending on the number of electrons present. For example, the expectation value of $\sigma_z$, in the ground state, is given by
\be
<0|\sigma_{z}|0>= <1|\sigma_{z}|1> = u^2(0)-v^2(0) = -g/\sqrt{1+g^2},
\label{sigz0}
\ee
and
\be
<2|\sigma_{z}|2> = u^2(2)-v^2(2) = +g/\sqrt{1+g^2}.
\label{sigz2}
\ee
Similarly, for the ground state, we obtain
\be
<0|\sigma_{x}|0>= <1|\sigma_{x}|1> = <2|\sigma_{x}|2> = - 1/\sqrt{1+g^2}.
\label{sigx012}
\ee

For large $g$ the $z$-component of the pseudospin switches from close to $-1$ to a value close to $+1$ as the occupancy changes from one to two electrons, but does not change when
the occupancy changes from zero to one electron. In contrast, the $x$-component remains constant as the occupancy changes. Note that this occurs independently of the value of the on-site Coulomb repulsion, $U$; in particular, the excitation
energy associated with an excited pseudospin state is given by the difference of Eqs. (\ref{eigenvalues}),
\be
\Omega_0 = 2 \omega_0 \sqrt{1 + g^2}.
\label{excitation}
\ee

\subsection{Mott transition}
In this subsection, we examine how the dynamics of the auxiliary boson field affects the Mott transition in the Dynamic Hubbard model. In the next subsection we
consider effects related to electron-hole asymmetry and `undressing'.

\begin{figure}[tp]
\begin{center}
\includegraphics[height=4.5in, width=3in]{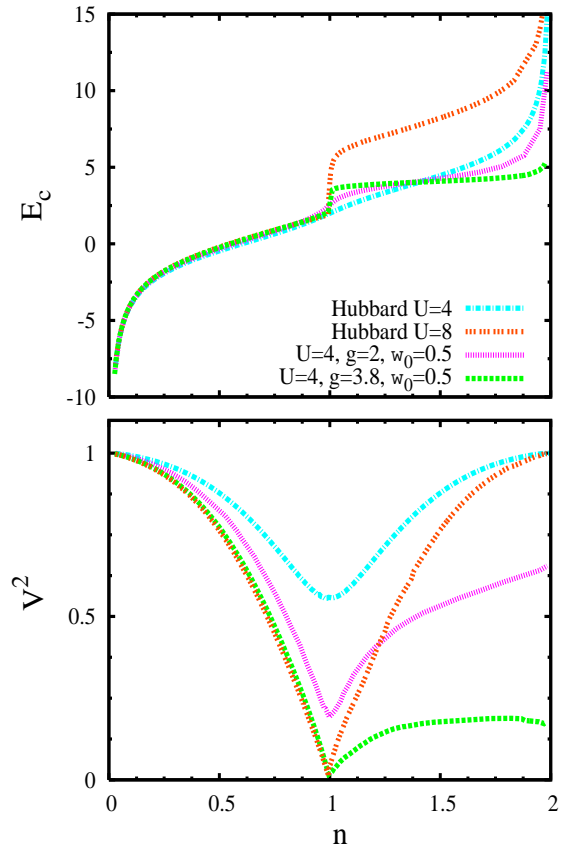}
\caption{Bath parameters (a) $E_{c}$ and (b) $V$ as a function of electron filling, for parameters with and without a Mott transition. For the static Hubbard model we use $U=8$ and $U=4$, as cases with and without a Mott gap, to illustrate the expected features in either case. The disappearance of $V^{2}$ at half-filling demonstrates the vanishing of quasiparticle spectral weight (see Eq. (\protect\ref{vself})), and the vertical increase of $E_c$ at half-filling is also a symptom of the Mott gap. Note, in the cases with non-zero $g$ (dynamic Hubbard model), the asymmetry with respect to half-filling.}
\end{center}
\end{figure}

The effect of the dynamic field on the Mott transition is best addressed near half-filling. One of the signatures of this transition, as approached from the Fermi Liquid side, is the disappearance of quasi-particle weight (QW). As noted in Eq. (\ref{vself}) and below, the self-consistent parameter $V$ measures the quasi-particle spectral weight. In Fig. 1 we show results for both self-consistent parameters, $E_c$ and $V$ for a number of parameters. Focusing on the static Hubbard model, it is clear that, within the DMFT approximation, a Mott transition takes place for a critical value of $4< U_c <8$, since, for $U=4$ the parameter $V$ remains non-zero over all electron fillings, while, for $U=8$ the parameter $V$ dips to zero at half-filling. Similarly, upon examining $E_c$ vs. $n$ (Fig. 1(a)), one sees a vertical jump at half-filling (present for $U=8$ but absent for $U=4$) as the characteristic signature of the Mott phase. The other two parameter choices illustrate that, for sufficiently large pseudospin coupling $g$, the Mott transition occurs, even for modest values of the bare Hubbard interaction, $U$. Thus the Mott transition is induced for $U=4$ with $g=3.8$, for example. This fact is further reinforced in Fig. 1(a), and redrawn in Fig. 2 using the relation between $E_c$ and $\mu$ that is implicit in Eq. (\ref{nself}), where now a plateau is present near half-filling for the parameter set $U=4$ and $g=3.8$, thus indicating the occurrence of a Mott transition.

\begin{figure}[tp]
\begin{center}
\includegraphics[height=2.5in,width=3in]{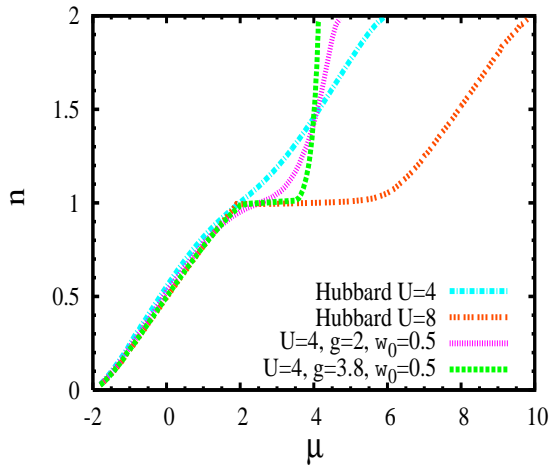}
\caption{(Color online)
Electron filling as a function of chemical potential, for the cases indicated in Fig. 1. Note the plateaus near half-filling, indicative of the Mott transition.}
\end{center}
\end{figure}

Further insight into the occurrence of the Mott insulator (and the inherent particle-hole asymmetry in this model described in the next subsection) can be gained by examining the behaviour of the expectation values of the pseudospin variables. We plot in Fig. 3 the expectation values $\langle \sigma_{x} \rangle $ and $\langle \sigma_{z} \rangle $ as a function of filling for two sets of parameters, one in which a metal-insulator transition does not take place ($g=2$, shown in pink), and one in which it does ($g=3.8$, shown in green). We should note that our values are in quantitative agreement with those of Ref. (\onlinecite{bouadim08}) (note that our values for $\langle \sigma_{x} \rangle $ are negative while their's are positive). Fig. 3 makes clear that below half-filling the pseudospin expectation values are fairly constant as a function of electron filling. An expectation value of $\langle \sigma_{z} \rangle = -1$
maintains an effective $U$ that is $U+2g\omega_0$, much higher than $U$ itself, so that double occupancy is restricted (see Fig. 5 below).

\begin{figure}[tp]
\begin{center}
\includegraphics[height=2.5in, width=3.0in]{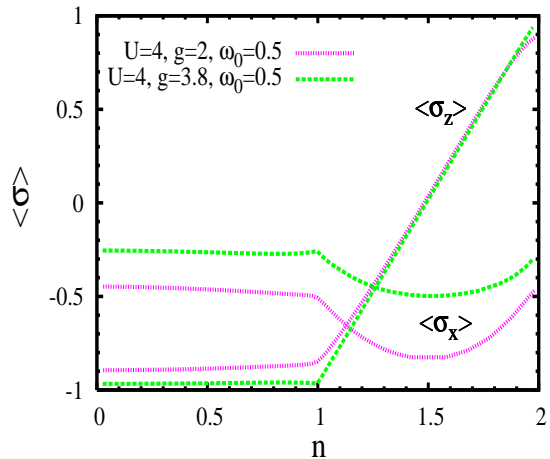}
\caption{(Color online)
The pseudospin expectation values $<\sigma_{x}>$ and $<\sigma_{z}>$ as a function of electron filling, for the dynamic Hubbard model, with $g=3.8$ (with a Mott gap transition), and $g=2$ (without a Mott transition). Note that the pseudospin plays very little role below half-filling, but undergoes a big change above half-filling.}
\end{center}
\end{figure}

A simple way to view the metal-insulator transition is through a variational approach, where, on the one hand, one uses the product state in which the electrons form a Fermi sea, while the pseudospins remain in their ground state, $\sigma_{zi} = -1$, for all sites, $i$. A simple calculation yields
\begin{equation}
{\langle H \rangle_{\rm Fermi} \over N} = -2tn(2-n) -g\omega_0 + (U+2g\omega_0)n^2/4,
\label{var1}
\end{equation}
where $n$ is the electron concentration, and we have used a simplifying assumption that the electron density of states is a constant, $g(\epsilon) = 1/(8t)$, appropriate to a 2D tight-binding model with nearest neighbour hopping $t$. This is essentially a Hartree calculation. The competing state is an insulator, with one electron per site (at least up to half-filling), no hopping, and a similar pseudospin state, $\sigma_{zi} = -1$, for all sites, $i$. The energy per site for this state is $-g\omega_0$. Therefore, restricting ourselves to half-filling, a metal-insulator transition will occur for $U$ beyond $U_{\rm crit}$, where $U_{\rm crit} = 8t - 2g\omega_0$. The critical value of $U$ is clearly lower as $g$ increases; this is because while the pseudospins remain in their ground state the effective value of on-site Coulomb repulsion is increased by the presence of the coupling to the pseudospin. Therefore $U$ itself can be smaller and, in combination with the effect of the pseudospin, still instigate a metal-insulator transition. Note that the values of $\langle \sigma_z \rangle$, as given by Eq. (\ref{sigz0}) for low filling are $-0.89$ and $-0.97$ for $g = 2$ and $g = 3.8$, respectively; these are close to $-1$, as used in the variational calculation, and also agree very well with the numerical results shown in Fig. 3. This in turn leads to a more accurate estimate of the {\em bare} on-site repulsion, given by \cite{hirsch02b}
\begin{equation}
U_{\rm bare} = U + {2g^2 \omega_0 \over \sqrt{1+g^2}},
\label{ubare}
\end{equation}
which results when the background degree of freedom is not allowed to relax.

As the filling increases above half-filling, the expectation values of the pseudospins change markedly. Electrons are no longer able to avoid double occupancy, so the pseudospin steadily changes from the $|->$ state to the $|+>$ state, to lower the on-site energy from $U_{\rm max} = U + 2g\omega_0 $ to $U_{\rm min} = U - 2g \omega_0$. As Fig. 3 indicates, there is essentially a linear increase of $<\sigma_{z}>$ from $-\frac{g}{\sqrt{1+g^2}} \sim -1$ to $+\frac{g}{\sqrt{1+g^2}} \sim +1$ as the band becomes completely full ($n=2$), as indicated by Eqs. (\ref{sigz0}) and (\ref{sigz2}). The relaxation of the pseudospin degree of freedom results in a lower quasiparticle weight, as Fig. 1(b) indicates. In Fig. 2, the curve representing $g=3.8$
(shown in green) approaches full occupation at $\mu \approx 4$. This point can be understood by the fact that $U_{\rm bare}$ is approximately zero when $n = 2$ (see Eq. (\ref{ubare})), so that the chemical potential goes to the top of the bare band ($2t^\ast$) plus the energy shift due to the pseudospin ${\omega_0 \over \sqrt{g^2 + 1}} (g^2 - 1) \approx 1.7t^\ast$.

\begin{figure}[tp]
\begin{center}
\includegraphics[height=6in, width=3in]{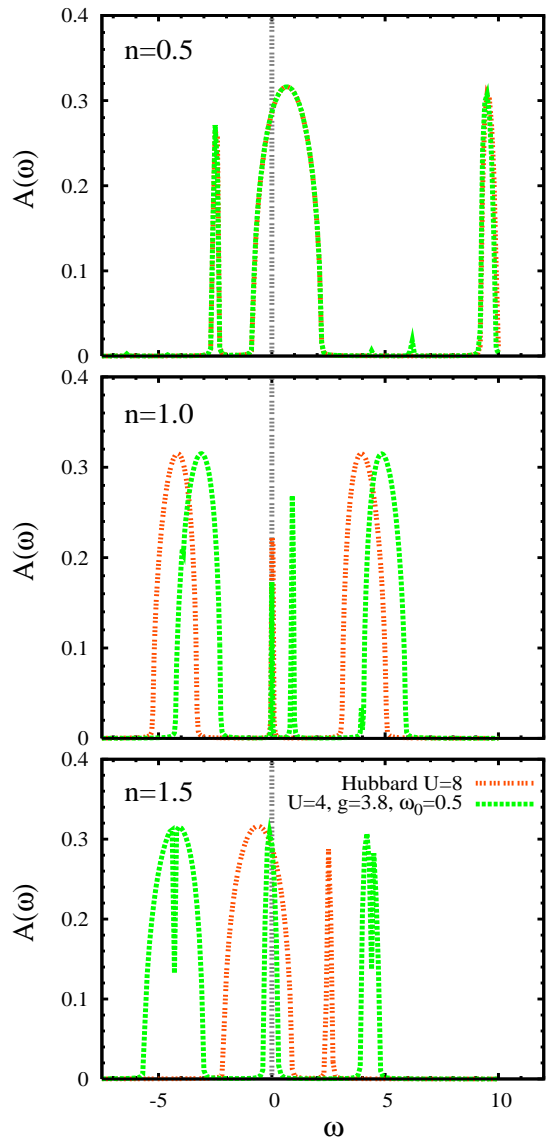}
\caption{(Color online)
A comparison of the spectral function for the static model with $U=8$ to the one for the dynamic model with  $U=4$, $g=3.8$, and $\omega_{0}=0.5$, for three different fillings. The two models behave very similarly below half-filling; at half-filling both undergo a Mott transition with the disappearance of the width of the resonance peak at zero frequency. Above half-filling the two models differ markedly; in particular, the peak width at the Fermi level is considerably smaller for the dynamic Hubbard model compared with the static Hubbard model.}
\end{center}
\end{figure}

As pointed out in Ref. (\onlinecite{potthoff01}), for the Hubbard model, the two-site DMFT approximation correctly produces three peaks in the single particle density of states corresponding to the lower and upper Hubbard bands, and a quasi-particle resonance peak at the Fermi energy, with quasi-particle weight $z$. For the dynamic Hubbard model, we show in Fig. 4 results for the parameters $U=4$, $g=3.8$ and $\omega_{0}=0.5$, along with results for the Hubbard model with $U=8$, for a number of different electron densities. Because of the pseudospin degree of freedom, the spectrum should contain at least four peaks, due to the appearance of more than two poles in the self energy; this is clearly the case in Fig. 4.
Below half-filling (Fig. 4(a)), the peak structure is very similar to that found in the Hubbard model, with $U=8$. At half-filling (Fig. 4(b), $n = 1.0$), the resonant peak at the Fermi level has all but disappeared, indicative of the Mott transition. Also shown are the results for much higher filling (Fig. 4(c), $n=1.5$), where the results are clearly not symmetric with those at $n=0.5$ (Fig. 4(a)), and certainly no longer similar to the results for the Hubbard model with $U=8$. As found in Ref. (\onlinecite{bouadim08}), the peak near the Fermi energy is considerably sharper at $n=1.5$ compared with $n=0.5$, indicating that this model is less free electron-like at high electron filling compared to low electron filling (as seen in Fig. 1(b) as well).

\begin{figure}[tp]
\begin{center}
\includegraphics[height=2.6in, width=3in]{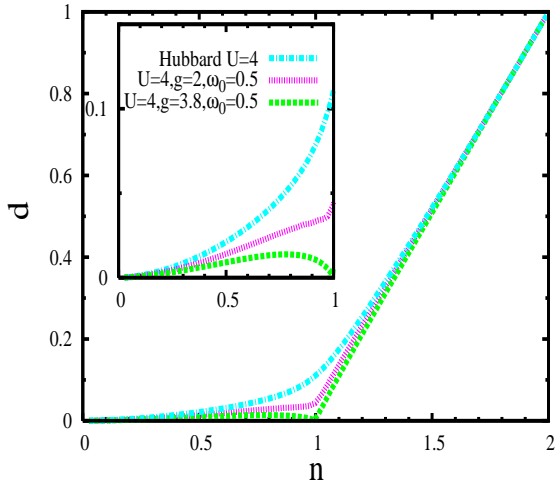}
\caption{(Color online)
The dependence of double occupancy vs. electron filling, for various parameters, for both the static and dynamic Hubbard models. For sufficiently large $g$ in the dynamic Hubbard model, the double occupancy is driven to zero at half-filling.}
\end{center}
\end{figure}

\begin{figure}[tp]
\begin{center}
\includegraphics[height=2.5in, width=3in]{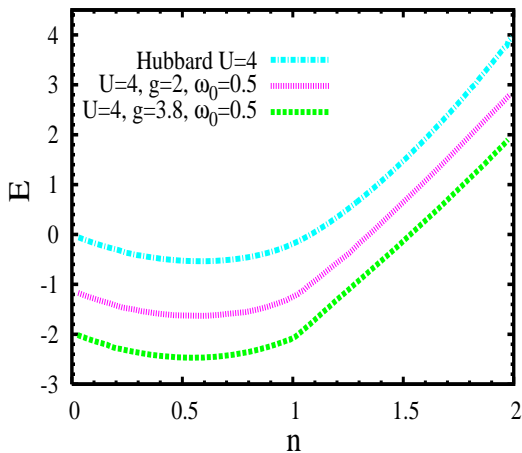}
\caption{(Color online)
The total ground state energy vs. electron filling for the same sets of parameters as in Fig. 5. The values at $n=0$ and $n=2$ agree with the analytically derived values, as explained in the text.}
\end{center}
\end{figure}

Fig. 5 shows the double occupancy as a function of electron filling. Again, very little double occupancy is present below half filling, as expected, though it is clear that the presence of the electron-pseudospin coupling $g$ suppresses the double occupancy near half-filling, and therefore enhances the Mott-like features of the Hubbard model. Above half-filling, the double occupancy quickly rises; though it is a more subtle effect here, the presence of $g$ enhances the double occupancy since, as the filling increases the effective Hubbard $U$ is decreasing due to the relaxation of the pseudospin degrees of freedom.

In Fig. 6 we show the total energy as a function of electron filling, for a number of parameters, as indicated in the figure. For an empty or completely full band the numerical results agree with those given by the analytical results obtained for the site Hamiltonian, Eq. (\ref{ham_onsite}):
\begin{eqnarray}
E(n=0) &=& -\omega_0 \sqrt{1+g^2} \\
E(n=2) &=& U -\omega_0 \sqrt{1+g^2}.
\label{energy_limits}
\end{eqnarray}
The results in Ref. (\onlinecite{bouadim08}) (their Fig. 14) are also in excellent agreement with the exact results given by the above equations.

\subsection{Electron-hole asymmetry and undressing phenomenology}

It was proposed in Ref. (\onlinecite{hirsch02c}) that the dynamic Hubbard model describes superconductivity driven by ``undressing'': namely, that when the Fermi level is near the
top of the band, pairing of hole carriers will  lead to transfer of   spectral weight from high to low frequencies and in particular in an increase of the
quasiparticle weight and a decrease in the effective mass. These effects should appear both in the single particle spectral function and in two particle spectral functions such as the
optical conductivity.

\begin{figure}[tp]
\begin{center}
\includegraphics[height=2.5in, width=3in]{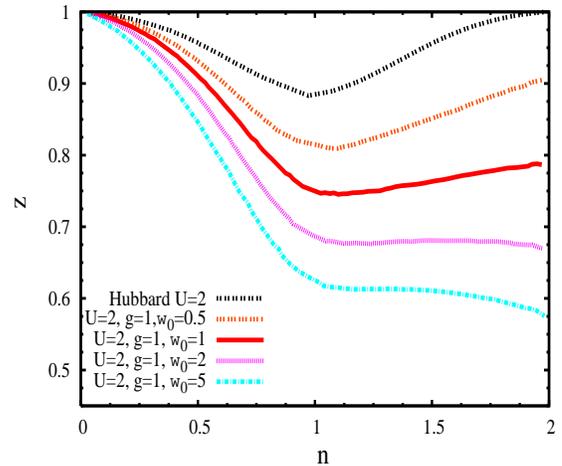}
\caption{(Color online)
The quasiparticle spectral weight, $z$ vs. electron filling, for a variety of parameter values. For the dynamic Hubbard model, increasing the pseudospin frequency $\omega_0$ leads to a steady decrease in the spectral weight, particularly at high value of electron filling, thus making the model more asymmetric with respect to half filling.}
\end{center}
\end{figure}

In the present paper we do not consider pairing correlation functions, and hence we cannot ascertain from our results whether or not the model describes superconductivity. However, we can
study properties of the model under hole doping. It is expected that the effects discussed in the previous paragraph should also occur both in the normal and the superconducting
state for an almost filled electron band
as a function of increased hole doping \cite{hirsch02c}.

\begin{figure}[tp]
\begin{center}
\includegraphics[height=2.5in, width=3in]{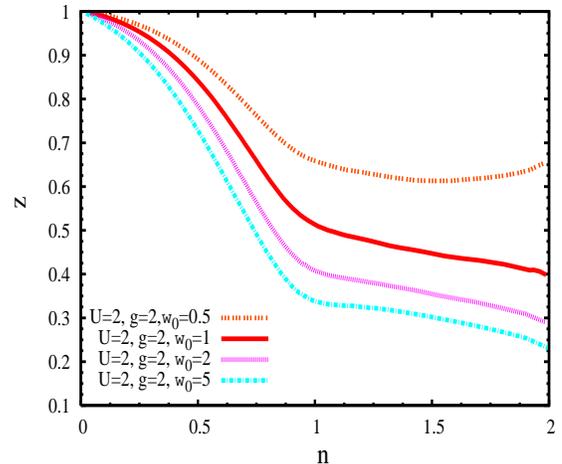}
\caption{(Color online)
Same as Fig. 7, except that now $g$ is increased to a value $g=2$. Increasing $g$ also increases the asymmetry with respect to half filling.}
\end{center}
\end{figure}

For the parameters considered in the previous subsection, it was found that doping the full band with holes led to a decrease rather than an increase in the
quasiparticle weight, in contradiction to these expectations. Thus we also do not expect superconductivity driven by ``undressing'' for those parameters. However, we find that the
expected ``undressing'' behavior does occur  for lower values of the on-site repulsion $U$ and/or larger values of the coupling $g$ as well as for higher values of the boson frequency $\omega_0$.

\begin{figure}[tp]
\begin{center}
\includegraphics[height=2.5in, width=3in]{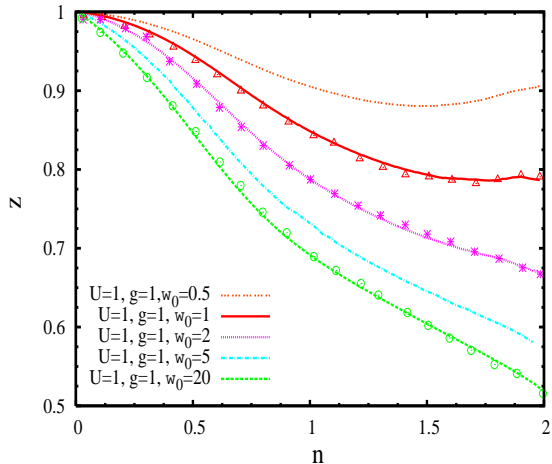}
\caption{(Color online)
Same as Fig. 7, but with a smaller value of $U$ (here $U=1$), and we have added a higher frequency result ($\omega_0 = 20$). With respect to Fig. 7, Mott physics is de-emphasized; instead, a decrease in spectral weight as electron filling increases arises primarily because of the role of the pseudospin degree of freedom. Note that the curves arise from data obtained through converging the parameter $V$ through the DMFT iterative process. The symbols come from integrating the spectral function peak at the Fermi level (see Fig. 12 below), and the good agreement is clear.}
\end{center}
\end{figure}

Figure 7 shows the behavior of the quasiparticle weight $z$ (recall $V = z^2$) for $U=2$, $g=1$  and various values of $\omega_0$ versus band filling.
It can be seen that for $\omega_0\ge 2$ the quasiparticle weight indeed increases
when the full band is doped with holes (i.e. the quasiparticle weight decreases with electron filling as $n \rightarrow 2$).
Fig. 8 shows the same behavior occurring already for $\omega_0\ge 1$ when we increase the value of the coupling to $g=2$ with $U=2$.
Similarly, the same behaviour occurring  for $\omega_0 \ge 1$ can be obtained by reducing the on-site repulsion to $U=1$ while keeping $g=1$ (Figure 9).
For these parameter ranges a Mott transition at half-filling does not occur.

\begin{figure}[tp]
\begin{center}
\includegraphics[height=2.5in, width=3in]{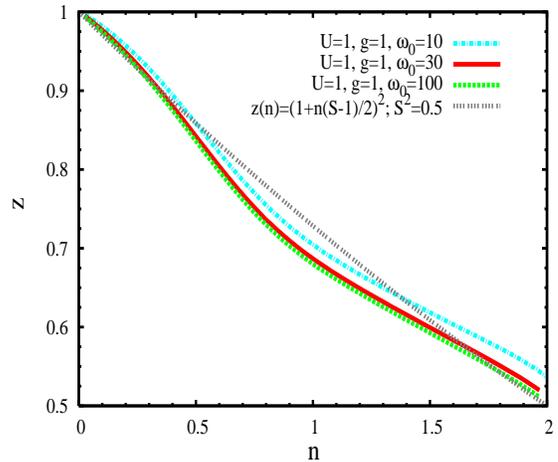}
\caption{(Color online)
Same as Fig. 9, except that now we have increased $\omega_0$ to very large values; in this limit we should recover the correlated hopping model, and the result at low hole doping (near full electron occupation) should follow the analytical result as indicated.\protect\cite{hirsch02c} This is clearly the case.}
\end{center}
\end{figure}

\begin{figure}[tp]
\begin{center}
\includegraphics[height=2.5in, width=3in]{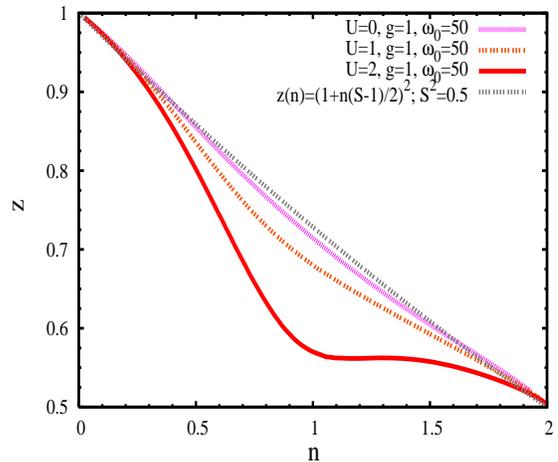}
\caption{(Color online)
To disentangle spectral weight reduction due to the Hubbard $U$ vs. the dynamic pseudospin effect ($g$), we show the quasiparticle spectral weight $z$ vs. electron filling $n$ for several values of $U$, including $U=0$. All numerical results shown are for a large value of pseudospin frequency, $\omega_0 = 50$. Agreement with the analytical result is excellent for $U=0$. Discrepancies for other values of $U$ shown come primarily from the quasiparticle reduction due to $U$, which is not accounted for in the analytical result.}
\end{center}
\end{figure}

As shown in Fig. 10, for large values of the frequency $\omega_0$, the behavior of the quasiparticle weight versus band filling $n$ is described approximately by the expression
\be
z=[1+(S-1)\frac{n}{2}]^2
\label{zanalytic}
\ee
with
\be
S^2=\frac{1}{1+g^2}
\label{sdefinition}
\ee
as expected.\cite{hirsch02c} Also, as Fig. 10 shows, for smaller values of $\omega_0$ the $n-$dependence of $z$ is qualitatively similar but, as $n \rightarrow 2$, the magnitude is larger than that given by Eq. (\ref{zanalytic}).
This dependence of quasiparticle weight on the boson frequency is consistent with the behavior found in Ref. \onlinecite{marsiglio03} using a
generalized Lang-Firsov transformation
within an Eliashberg treatment. It was also found in that work that smaller boson frequency enhances the tendency to pairing.

\begin{figure}[tp]
\begin{center}
\includegraphics[height=7.5in, width=3.3in]{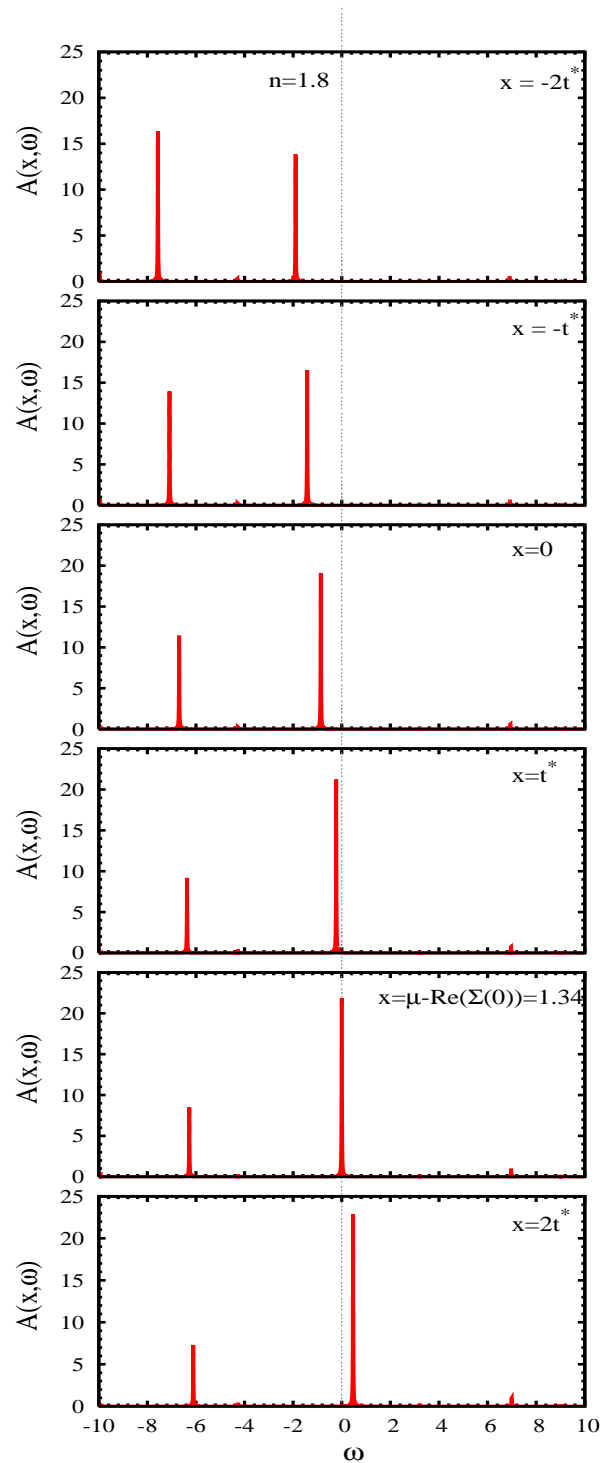}
\caption{(Color online)
The spectral function for various energies, as shown, for $n=1.8$. Here, we have used $U=1$, $g=1$, and $\omega_0 = 2$ (see pink curve in Fig. 9). The spectral functions consist primarily of two peaks, separated by an energy corresponding to the pseudospin excitation energy, $\Omega_0$. The 2nd last frame shows the spectral function at an energy corresponding to the Fermi level, and the weight under the peak at $\omega = 0$ corresponds to the quasiparticle residue, $z$.}
\end{center}
\end{figure}

Figure 11 shows that the analytical form, Eq. (\ref{zanalytic}) is indeed very accurate when $U$ is not present. Results shown for increasing values of $U$ indicate {\em additional} decreasing of quasiparticle weight that occurs due to well documented `Hubbard physics'. Nonetheless, for all cases, the overall decreasing trend as a function of electron filling is clearly coming from `quasiparticle dressing' due to the pseudospin degree of freedom (Eqs. (\ref{zanalytic}) and (\ref{sdefinition})).

\begin{figure}[tp]
\begin{center}
\includegraphics[height=7in, width=3.3in]{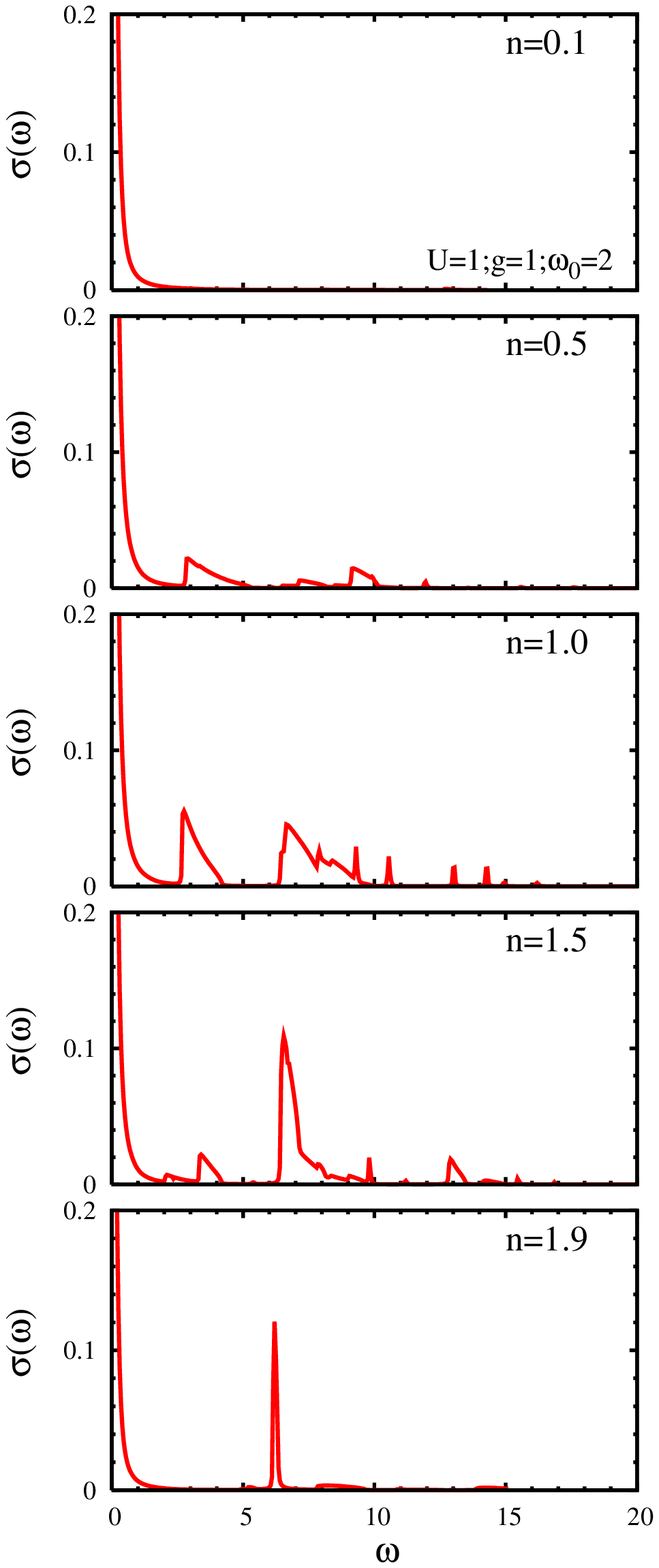}
\caption{(Color online)
The optical conductivity is shown at different filling for $U=1,g=1,\omega_0=2$. Though not so apparent in this figure, there are four primary components with characteristic frequencies, $\omega=0$ (Drude), $\omega_L \approx  U/2 + \sqrt{(U/2)^2 + 4t^2S^2} \approx 2$, $\omega_M=\Omega_0=2\omega_0 \sqrt{1+g^2} \approx 5.7$ and $\omega_H = 2\Omega_0 \approx 11.3$.  }
\end{center}
\end{figure}

In Figure 12 we show the behavior of the single particle spectral function for one particular electron filling, $n=1.8$, for a variety of 'momenta'. Each spectral function consists primarily of two peaks, separated by roughly the pseudospin excitation energy $\Omega_0$ given by Eq. (\ref{excitation}). The weight of the quasiparticle peak at the Fermi level, shown in the 5th panel, corresponds to the residue $z$ plotted for all fillings in Fig. 9 (middle curve). As shown there, this quasiparticle weight decreases (increases) with increasing electron (hole) concentration. In Fig. 13 we show the behaviour of the optical conductivity obtained using Eq. (\ref{cond}), for several electron fillings.
An exact calculation of the optical conductivity for a two-site model \cite{Hi} shows that the optical conductivity can generally be divided into four contributions, as we now describe.

\begin{figure}[tp]
\begin{center}
\includegraphics[height=2.5in, width=3in]{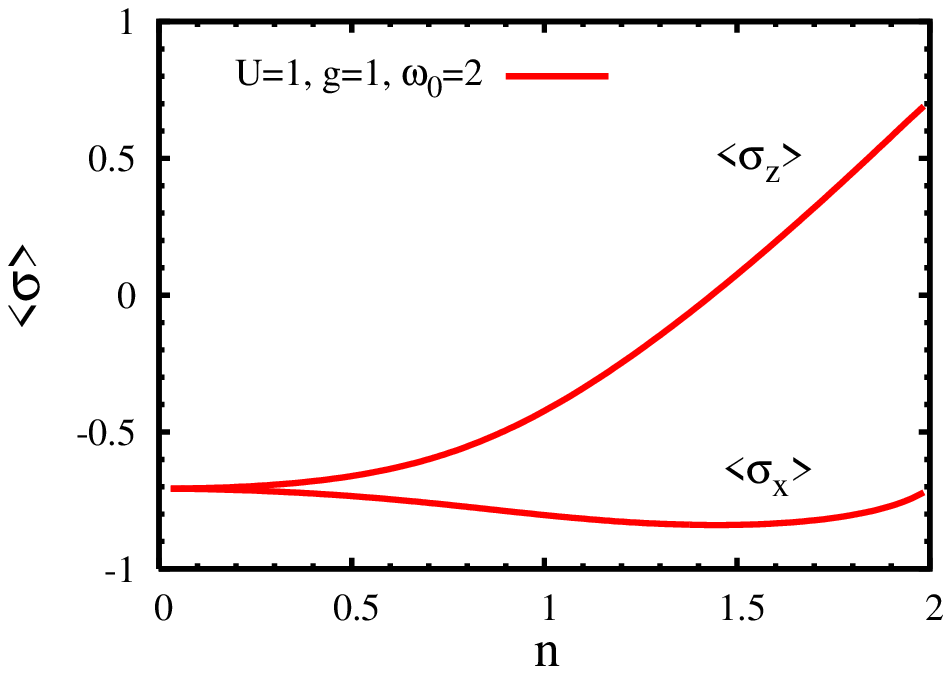}
\caption{(Color online)
The pseudospin expectation values $<\sigma_x>$ and $<\sigma_z>$ as a function of electron filling with $U=1,g=1,\omega_0 = 2$ }
\end{center}
\end{figure}

There will be a Drude part centered at $\omega = 0$ with a width $1/\tau$, normally due to elastic impurity scattering.
In our calculations, this part appears as a $\delta$-function at the origin, with artificially imposed broadening (see Fig. 13). This component involves transitions between the two coherent parts of the spectral functions (see Eq. (\ref{cond})). For low electron fillings these transitions give rise to essentially the entire frequency dependent conductivity.

\begin{figure}[tp]
\begin{center}
\includegraphics[height=2.5in, width=3in]{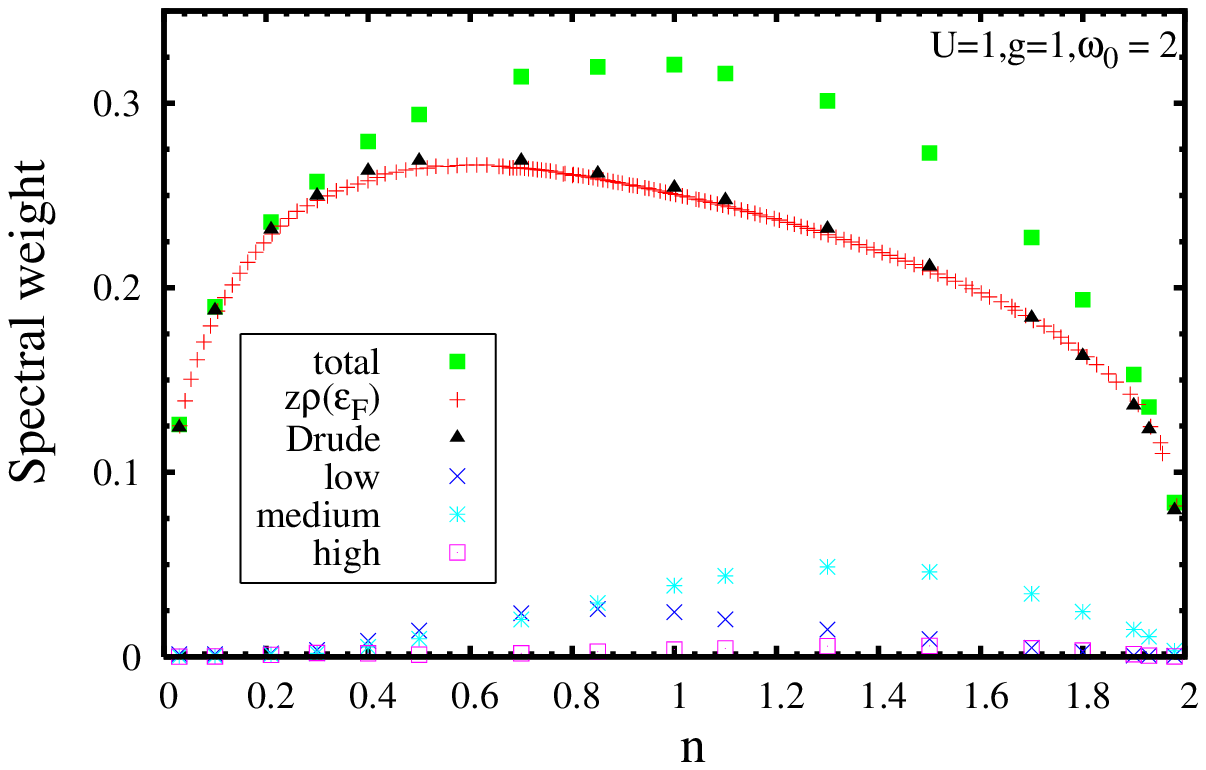}
\caption{(Color online)
The spectral weight of the optical conductivity, as a function of electron filling. The total weight, along with the weight of each component (see figure caption in Fig. 13), is plotted as a function of electron filling. As in Fig. 13, we have used the same parameter set $U=1,g=1,\omega_0=2.0$.}
\end{center}
\end{figure}

Three more components contribute as the electron filling comes close to and exceeds half-filling. They have characteristic frequencies $\omega_i$, and weights $C_i$, for $i=L,M,H$, corresponding to 'low', 'medium', and 'high'. These designations are relevant only when $U$ is sufficiently small, as in Fig. 13. Then the 'low' frequency part is peaked near $\omega_L \approx  U/2 + \sqrt{(U/2)^2 + 4t^2S^2}$, and involves transitions between the ground state and an excited state with a doubly occupied site (both states have pseudospins in their ground states). The quantity $S \equiv \langle 1 | 2 \rangle$
gives an estimate of the polaronic effect of the pseudospin excitation required as an electron undergoes a hop.\cite{Hi}

The characteristic frequency for the 'medium' range is $\omega_M \approx \Omega_0 = 2\omega_0\sqrt{1+g^2}$. This is the excitation energy for a pseudospin (see Eq. (\ref{excitation})), and transitions between states differing by a pseudospin excitation are responsible for this part of the conductivity. These transitions will play a more significant role as the electron-pseudospin coupling strength, $g$, increases.

Finally, the high frequency characteristic frequency is $\omega_H \approx 2\Omega_0$, and requires a transition in which two pseudospin excitations are created. This necessarily tends to happen only when the electron concentration is very high, i.e. in the hole region. The weights of these various contributions are difficult to estimate in advance; they will depend on the hopping overlap integrals. Fig. 13 shows the optical conductivity at various fillings for $U=1$, $g=1$, and $\omega_0=2.0$; then $\omega_L \approx 2.0$, $\omega_M \approx 5.7$ and $\omega_H \approx 11.3$. For these same microscopic parameters, we also show the expectation values of the $z$ and $x$ components of the pseudospin degree of freedom as a function of electron density in Fig. 14. Their values are given analytically at $n=0$ and $n=2$: $\langle \sigma_x \rangle = -1/\sqrt{1 + g^2}$ for both limits and $\langle \sigma_z \rangle = -g/\sqrt{1 + g^2}$ ($g/\sqrt{1 + g^2}$) for $n=0$ ($n=2$), and DMFT obviously gets these correctly. The steady increase of $\langle \sigma_z \rangle$ with increasing electron density reflects the increased occurrence of the excited pseudospin state in the ground state as the lattice becomes more crowded.

A more in-depth understanding of the optical conductivity comes from examining the spectral weight, partitioned into the various contributions, as described above. For concreteness we define the Drude conductivity to include contributions in the range, $0 < \omega < \omega_L$. The other contributions are defined as follows: low frequency in the range $\omega_L < \omega < \omega_M$, midrange for $\omega_M < \omega < \omega_H$, and the high frequency range for $\omega > \omega_H$. This is shown in Fig. 15. First, note that the total spectral weight is generally asymmetric as a function of electron concentration. Note, moreover, that as the electron concentration approaches zero or full filling, the spectral weight approaches zero, since there are no carriers in either case. The Drude weight, defined as the area under the near-zero frequency portion of the conductivity, is also asymmetric; this is shown by the black triangular points. An integration over the Drude portion of Eq. (\ref{cond}) shows that the Drude spectral weight satisfies the sum rule \cite{georges96}:
\be
\int_0^ \infty \sigma_{\rm Dr}(\omega)d\omega=- \frac{\pi e^2a^2}{2d \hbar^2 \nu} <K> = \frac{{\omega_p^*}^2}{8\pi}
\ee
and in the limit of infinite dimension $d \rightarrow \infty$:
\be
\frac{{\omega_p^*}^2}{4\pi} = \frac{4\pi t^2 e^2 a^2}{\hbar^2 \nu} z \rho(\epsilon_F)
\label{drude_weight}
\ee
Here $\omega_p^*$ is the renormalized plasma frequency. We can see from Eq. (\ref{drude_weight}) that the Drude weight depends on the quasiparticle weight $z$ and the DOS at $\epsilon_F=\mu-Re(\Sigma(0))$. The result from Eq. (\ref{drude_weight}) is also plotted in Fig. 15, and is even more asymmetric than the total spectral weight. In addition, other contributions are also plotted as a function of electron filling.

Note that the total weight is Drude-like both for $n \rightarrow 0$ and for $n \rightarrow 2$, i.e. the contributions from the finite frequency portions fall off more quickly. Nonetheless, it is clear that spectral weight at high frequency is most intense in the hole-like region ($n > 1$) as compared with the electron-like region ($n < 1$). This can be understood by the following argument. First, we focus on the (minor) role of transitions involving the Hubbard $U$. These are described by the so-called 'low' frequency contribution, and they peak at half-filling. This is because that filling corresponds to the situation where more fluctuations are liable to occur. At low fillings $U$ hardly plays any role, while at high filling $U$, by virtue of playing the same role for almost all electrons, again hardly plays any role.

With excitations involving the pseudospin degree of freedom the situation is a little more subtle. At low to intermediate fillings, the pseudospins play almost no role because single electron transitions do not require a relaxation of the pseudospin degree of freedom. At high to intermediate fillings, however, transitions generally involve the pseudospin degree of freedom, because the overlap between a singly occupied state and a doubly occupied state, $S \equiv \langle 1 | 2 \rangle$, can be much less than unity (in contrast, for this model, the corresponding overlap between an empty and singly occupied site, $T \equiv \langle 0 | 1\rangle$ remains unity). Thus, there is an electron-hole asymmetry, with the pseudospin physics playing a large role in the electron density region $1 < n < 2$. In fact, variation in the pseudospin expectation value, $\langle \sigma_x \rangle$, reflects this fact also. Note the resemblance between the electron density dependence of $\langle \sigma_x \rangle$ in Fig. 14 (inverted) with the electron density dependence shown by the spectral weight in the medium frequency range shown by the blue asterisks in Fig. 15 (the parameter regimes used in Fig. 3 depict this variation of $\langle \sigma_x \rangle$ in the hole-doped region even more pronouncedly). Hence, optical spectral weight in the medium frequency region plays a more important role for electron densities on the hole-like side of the phase diagram.
This is consistent with the explanation in Ref. (\onlinecite{bouadim08}), that since $\langle \sigma_x \rangle$ measures the fluctuations of $\sigma_z$ (in imaginary time), it should achieve a minimum (they obtained a maximum for reasons that are unclear) at the point where $\langle \sigma_z \rangle$ is changing the most quickly (as a function of electron density). This tends to occur midway through the hole-like side of the phase diagram, i.e. near $n \approx 1.5$.

Such transfer of spectral weight from high to low frequencies as function of increasing hole concentration is seen in high $T_c$ cuprates, both in the single particle spectral function (photoemission experiments)\cite{photo01} and in the two particle spectral function (optical conductivity)
\cite{basov99,santander01,marel02,carbone06}.

\section{Conclusions}

We have investigated various properties of the dynamic Hubbard model by using dynamical mean field theory in the two-site approximation. Where comparison was possible our results agree surprisingly well with Quantum Monte Carlo results \cite{bouadim08} performed on small clusters ($6\times 6$) at finite temperature ($\beta =5$).
In agreement with Ref. (\onlinecite{bouadim08}) we find that the presence of an auxiliary degree of freedom enhances the Mott transition at half filling. We see this in the vanishing of the quasiparticle spectral weight, in the appearance of a plateau in the $n$ vs. $\mu$ curve, and in the vanishing width of a quasiparticle peak in the density of states.

An important property of this model is its asymmetry with respect to half filling. The hole side ($n>1$) is always considerably more dressed than the electron side, because of the relaxation of the pseudospin degree of freedom. This occurs because electrons minimize their Coulomb repulsion, at a cost of becoming somewhat more sluggish in their movements, i.e. they form polaronic-like states.

We have identified a parameter regime where the electron-hole asymmetry is very clear (see Figs. 7-11), and the quasiparticle spectral weight increases linearly with hole doping away from $n=2$, as described in an effective model with correlated hopping.\cite{hirsch90} This quasiparticle undressing is a general phenomenon that occurs not only as a function of hole doping, but as a function of, for example, temperature changes and phase transitions.\cite{His} Understanding the degree to which this undressing is robust as the auxiliary degree of freedom is reduced from the anti-adiabatic limit to a more physical regime is one of the goals of this paper. Fig. 9 in particular illustrates that the degree of dressing (at $n=2$) is reduced as $\omega_0$ is reduced, and therefore the degree of `undressing' as the electron occupation is decreased from $n=2$ is reduced. Future work will determine the impact of this frequency scale on superconductivity.

We computed single particle spectral weights and the optical conductivity for this model. In both cases the frequency dependence is distinctly different for electron-like and hole-like doping levels. The asymmetric behavior for holes vs. electrons is clear as a function of doping; determining this asymmetry for a given doping level can be established through photoemission or tunneling experiments. \cite{marsiglio89}

\bibliographystyle{prb}

\end{document}